Enclosure: balloonsnn  [16,421 bytes]
----------------------------------------------------------------------
\documentstyle[12pt]{article}
\newlength{\extraspace}
\newlength{\extraspaces}
\begin{document}
\pagestyle{empty}
\begin{titlepage}
\begin{flushright}
UTPT-93-18
\\ hep-ph/9307359
\\ revised November 1993
\end{flushright}
\vspace{2.5cm}
\begin{center}
{\LARGE Cosmic balloons}\\ \vspace{40pt}
{\large B. Holdom\footnote{holdom@utcc.utoronto.ca}}
\vspace{0.5cm}

{\it Department of Physics\\ University of Toronto\\
Toronto, Ontario\\Canada M5S 1A7}
\vskip 2.1cm
\rm
\vspace{25pt}
{\bf ABSTRACT}

\vspace{12pt}
\baselineskip=18pt
\begin{minipage}{5in}

Cosmic balloons, consisting of relativistic particles trapped inside a
spherical domain wall, may be created in the early universe.  We
calculate the balloon mass $M$ as a function of the radius $R$ and the
energy density profile, $\rho (r)$, including the effects of gravity.
At the maximum balloon mass $2GM/R\approx 0.52$ for any value of the
mass density of the wall.

\end{minipage}
\end{center}
\vfill
\end{titlepage}
\pagebreak
\baselineskip=18pt
\pagestyle{plain}
\setcounter{page}{1}

Domain walls may form in the early universe during a phase transition
involving the spontaneous breaking of a discrete symmetry.  If the mass
density (or ``surface tension") of the wall, $\sigma $, is characterized by
any of the known scales of particle physics then the walls would quickly
come to dominate the energy density of the universe.  The infinite walls
which stretch across the horizon must then disappear before they dominate.
Possible mechanisms include small symmetry breaking effects which slightly
prefer one of the discrete number of vacua, or cosmic strings which can
provide the infinite walls with holes and edges.

Of interest here is the situation in which the discrete symmetry exchanges a
pair, or pairs, of particles. In this case a particle may have a different
mass on either side of the wall.  When a particle has total energy less than
the rest mass it would have on the other side of the wall then it can become
trapped inside a region having a closed domain wall as a boundary.
Collections of such particles so trapped form cosmic balloons.  Balloons of
irregular shape will oscillate and will eventually settle down to spheres
(possibly rotating) due to the emission of gravitational radiation (at the
very least).  A stable object forms in which the pressure of the gas inside
balances the surface tension of the wall.

An example of cosmic balloons which have been studied in some detail are
neutrino balls \cite{a}.  Here the discrete symmetry is parity, present in a
left-right symmetric theory of electroweak interactions.  The spontaneous
breaking of this discrete symmetry typically produces large (small) Majorana
masses for the right-(left)-handed neutrinos.  These masses are reversed on
the other side of a domain wall.  Neutrino balls are collections of trapped
light right-handed neutrinos.  For long-lived balls to form the initial
temperature of the neutrinos must be below $0.21{m}_{e}$, in which case the
neutrinos cool to a degenerate Fermi gas.  This occurs through the reaction
$\nu \overline{\nu }\rightarrow {e}^{+}{e}^{-}$, which then shuts off once
degeneracy is reached.  The much slower reaction $\nu \overline{\nu
}\rightarrow 3\gamma $ results in a lifetime of neutrinos balls of order the
age of the universe or longer.  To avoid the reaction $\nu \overline{\nu
}\rightarrow {e}^{+}{e}^{-}$ the chemical potential of the degenerate
neutrino gas must be less than ${m}_{e}$, and this in turn places a lower
bound on the radius and mass of a neutrino ball.

Neutrino balls may be of interest in several ways.  They are in a
supermassive mass range making them suitable for the seeding of galaxy
formation.  In an accretion mode they have been discussed in connection with
quasars, with the reaction $\nu \overline{\nu }e\rightarrow e\gamma $
possibly playing a role \cite{b}.  Stars passing into the neutrino ball will
be ``cooled", thus enhancing the probability of supernova type events within
the ball.  Neutrinos from such an event interact with the ambient neutrinos
in the ball and produce ${e}^{+}{e}^{-}$ pairs; the observational signal
would be a gamma ray burst \cite{c}.  The neutrino ball remains basically
unscathed by such an event, and it remains a site for further such events.

In this paper we shall be more general and define a cosmic balloon to be any
spherical domain wall trapping relativistic particles.  We wish to extract
the balloon mass $M$ as a function of its radius $R$ including the effects
of gravity.\footnote{We do not agree with the results in \cite{f}.}  This
function will be completely determined once the mass density of the wall,
$\sigma$, is given.  In the neutrino ball case ${\sigma}^{1/3}$ is
constrained to be less than a few TeV for long-lived balls to form \cite{a},
and greater than a few hundred GeV, since ${\sigma}^{1/3}$ is roughly
associated with the scale of left-right symmetry breaking.  For other types
of cosmic balloons we may allow $\sigma$ to be anything.  We shall restrict
our attention to static, nonrotating, balloons where all quantities are
functions of the distance $r$ from the center of the balloon.

Since we only treat a relativistic gas inside the balloon we have the simple
equation of state \begin{equation} p(r)={\frac{\rho
(r)}{3}}.\label{a}\end{equation}  The fundamental differential equation
deduced from Einstein equations, combined with (\ref{a}), reads \cite{d}
\begin{equation} -{r}^{2}\rho '(r)=4G{\cal M}(r)\rho (r)\left[{1+
{\frac{4\pi {r}^{3}\rho (r)}{3{\cal M}(r)}}}\right]{\left[{1- {\frac{2G{\cal
M}(r)}{r}}}\right]}^{-1},\label{d}\end{equation} where \begin{equation}
{\cal M}'(r)=4\pi {r}^{2}\rho (r)~~~{\rm and}~~~{\cal
M}(0)=0.\label{e}\end{equation}

The energy density of the gas close to the wall, $\rho (R)$, is fixed by the
requirement that the pressure $p(R)$ be sufficient to hold the wall up.  The
other forces on the wall are inward pointing, and are due to the surface
tension of the wall, the self-gravitational force of the wall on itself, and
the gravitational force due to the gas in the interior.  To find the required
$p(R)$ we may use the Gauss-Codazzi formalism; this was used in \cite{g} and
\cite{h} to find the motion of spherical dust walls and domain walls,
respectively, with empty space inside and outside the spherical shell.  We
will not give the details of this formalism here, but it is fairly
straightforward to extend it to our case to find the {\em static} solution
with the metric appropriate to the gas in the interior.  The latter takes
the form \begin{equation}d{s}^{2}=-B(r){dt}^{2}+A(r)d{r}^{2}+{r}^{2}(d{\theta
}^{2}+{\sin}^{2}\theta d{\phi }^{2})\:\:\:\:\:r<R\end{equation}  We have
\begin{equation}A(R)=\alpha^{-2},
\:\:\:\:\alpha^2=1-{\frac{2G{\cal M}(R)}{R}}.\label{w1}\end{equation}  ${\cal
M}(R)$ is the ``mass of the gas" as defined in (\ref{e}).

The equation for hydrostatic equilibrium \cite{d}, $B'/B=-2p'/(p+\rho )$,
combined with (\ref{a}) and (\ref{d}) yields
\begin{equation}\frac{B'(R)}{B(R)}={\frac{2G{\cal
M}(R)}{\alpha^2{R}^{2}}}\left({1+{\frac{4}{3}}\pi {R}^{3}{\frac{\rho
(R)}{{\cal M}(R)}}}\right).\label{w2}\end{equation}  Using (\ref{w1}) and
(\ref{w2}) in a treatment similar to
\cite{h} we find that for a static solution, $\rho(R)$ must satisfy
\begin{equation} \frac{\rho (R)}{3}={\frac{(\alpha +\beta )\sigma }{R}}-2\pi
G\sigma^2+{\frac{\sigma }{\alpha }}{\frac{G{\cal
M}(R)}{{R}^{2}}}\left({1+{\frac{4}{3}}\pi {R}^{3}{\frac{\rho (R)}{{\cal
M}(R)}}}\right)\label{b}\end{equation} with \begin{equation}{\beta
}^{2}=1-{\frac{2GM}{R}}.\end{equation}  $M$ is the total mass of the
balloon.  In (\ref{b}) we see that the pressure is balanced by the surface
tension, the self-gravitational attraction of the wall, and the
gravitational force due the gas in the interior.  The $1/\alpha$ factor in
the last term is a familiar factor which appears in the acceleration induced
by a Schwarzschild metric.  For the total mass of the balloon we find
\begin{equation} \frac{M}{\beta}={\frac{{\cal M}(R) }{\alpha
}}\left({1+{\frac{4}{3}}\pi {R}^{3}{\frac{\rho (R)}{{\cal
M}(R)}}}\right)-4\pi {R}^{2}\sigma .\label{c}\end{equation}

The first thing we notice about equations (\ref{d}), (\ref{e})
(\ref{b}), and (\ref{c}) is that they are invariant under the following
scaling transformation.  $$\rho (r)\rightarrow {\lambda
}^{-2}\rho
\left({{\frac{r}{\lambda }}}\right)~~,~~{\cal M}(r)\rightarrow \lambda {\cal
M}\left({{\frac{r}{\lambda }}}\right)~~,$$ \begin{equation}r\rightarrow
\lambda r~~,~~R\rightarrow \lambda
R~~,~~M\rightarrow \lambda M~~,~~\sigma
\rightarrow {\lambda }^{-1} \sigma.\label{f}\end{equation}  This implies that
once we have a set of solutions for a given $\sigma$, then we can find the
solutions for any other value of $\sigma$ via a simple scaling.

We first solve equations (\ref{d}) and (\ref{e}) for a particular value of
$\rho (0)\equiv e$.  This defines the function ${\rho }_{e}(r)$.  The radius
$R$ is then determined by the point at which (\ref{b}) is satisfied, if such
a point exists. These equations can of course be solved numerically.  But we
have also been able to find a good analytical approximation to the function
${\rho }_{e}(r)$.  We first find a Taylor series solution for ${\rho
}_{e}(r)$ and then convert it into the corresponding Pad\'{e} approximant.
The result from {\em Maple} \cite{e} is \begin{equation}{\rho }_{e}^{\rm
Pad\acute{e}}(r)=N/D.\end{equation}\begin{eqnarray}N&=&119350556325e-
18913737150\pi Ge^{2}r^{2}+\nonumber\\&&166307621760 \pi
^{2}G^{2}e^{3}r^{4}-85521303936\pi ^{3}G^{3}e^{4}r^{6}\end{eqnarray}
\begin{eqnarray}D&=&119350556325+617622563250\pi
Ger^{2}+\nonumber\\&&1253635451040\pi ^{2}G^{2}e ^{2}r^{4}+1001139948544\pi
^{3}G^{3}e^{3}r^{6}\end{eqnarray} The error is greatest when the effect of
gravity is the greatest, that is for the largest mass balloons and for
$r\approx R$.  In this case the error is less than a percent; in other cases
the deviation of ${\rho }_{e}^{\rm Pad\acute{e}}(r)$ from the numerical
solution is much less.  We also see that the scaling behavior is obeyed.
{}From the solution ${\rho }_{e}(r)$ another solution is ${\rho
}_{e/{\lambda}^2 }(r)={\lambda }^{-2}{\rho }_{e}\left({r/\lambda}\right)$.

Two examples of $M$ as a function of $R$ are given in Fig. (1) for the cases
$\sigma =1$ TeV${}^3$ and $\sigma =1/2$ TeV${}^3$.  For each point
$({R}_{1},{M}_{1})$ on the first curve there is a point
$({R}_{1/2},{M}_{1/2})=(2{R}_{1},2{M}_{1})$ on the second curve.  For the
smallest balloons gravity has little effect; the energy density is
essentially independent of $r$ and is given by (\ref{b}).  The resulting
mass from (\ref{c}) is approximately $12\pi {R}^{2}\sigma $; the latter is
the mass function in \cite{a} derived in the absence of gravity, shown by
the dotted curve in Fig. (1).  As the size of the balloon increases, the
energy density of the gas decreases due to (\ref{b}).  But at the same time
gravity has more effect, causing the energy density near the center of the
balloon to be greater than near the wall.  Corresponding to the points
labeled $A$ in Fig. (1) we plot $\rho (r)$ in Fig. (2).  As the size
continues to increase this tendency of the energy density to pile up at the
center of the balloon gradually overwhelms the general tendency of the
energy density to decrease.  $\rho (0)$ actually reaches a minimum at the
points labeled $B$, and we plot the corresponding $\rho (r)$ in Fig. (2).

$\rho (0)$ then starts to increase as both the $M$ and $R$ continue to
increase.  As gravity becomes even stronger, $R$ itself reaches a maximum as
$M$ continues to increase.  Then for increasing $M$, $R$ decreases.
Eventually a point is reached, labeled by $C$, at which $M$ reaches a
maximum; we again plot the corresponding $\rho (r)$ in Fig. (2).  At this
point \begin{equation}{\frac{\partial M}{\partial \rho
(0)}}=0.\end{equation}  This condition signals the point at which the
balloon passes from stability to instability \cite{d}.  Any balloon beyond
this point presumably collapses into a black hole.

Thus we have found a maximum mass for balloons which is determined solely by
the mass density of the wall:  \begin{equation}{M}_{{\rm
\max}}\approx{2.45\times 10}^{7}{M}_{\rm solar}{\left[{{\frac{\sigma }{\rm
Te{V}^{3}}}}\right]}^{-1}.\end{equation}  The associated balloon radius is
\begin{equation}{R}_{{M}_{{\rm \max}}}\approx{1.39\times 10}^{13}{\rm
cm}{\left[{{\frac{\sigma }{\rm Te{V}^{3}}}}\right]}^{-1}.\end{equation}
This is about 3\% smaller than the maximum radius, which occurs for slightly
smaller $M$. For any value of $\sigma$ we have \begin{equation}{\frac{2
G{M}_{{\rm \max}}}{{R}_{{M}_{{\rm \max}}}}}\approx 0.52.\end{equation}  This
relation is indicated by the heavy straight line in Fig. (1).  For
comparison $2GM/R=1$ is indicated by the dashed line.

Thus the maximum fractional red shift of a spectral line emitted in or close
to an isolated cosmic balloon, for any $\sigma$, is \begin{equation} z\equiv
{\frac{\Delta \lambda }{\lambda }}={\left({1-{\frac{2 G{M}_{{\rm
\max}}}{{R}_{{M}_{{\rm \max}}}}}}\right)}^{-1/2}-1\approx 0.45.\end{equation}

Perhaps the most intriguing property of cosmic balloons is that one finds a
connection between an interesting mass scale for particle physics, e.g. 1
$\mbox{TeV}^3$ for the mass density of the domain wall, and an interesting
mass scale for astrophysics, e.g. $10^7$ solar masses for the mass of the
balloon.  An open problem in astrophysics is the mechanism for the formation
of black holes in this mass range, which are needed to provide the
gravitational potential in accretion models of quasars.  Cosmic balloons may
replace black holes in this context and/or end up forming supermassive black
holes through accretion.  The results here provide a starting point for a
more detailed study of this possibility.

\vspace{2ex}
\noindent{\large Acknowledgments}

I would like to thank the Aspen Center for Physics, where this work was
completed.  I would like to thank N. Cornish for helping to correct an
error.  This research was supported in part by the Natural Sciences and
Engineering Research Council of Canada.
 \newpage \noindent{\bf Figure Captions} \vspace{3ex}

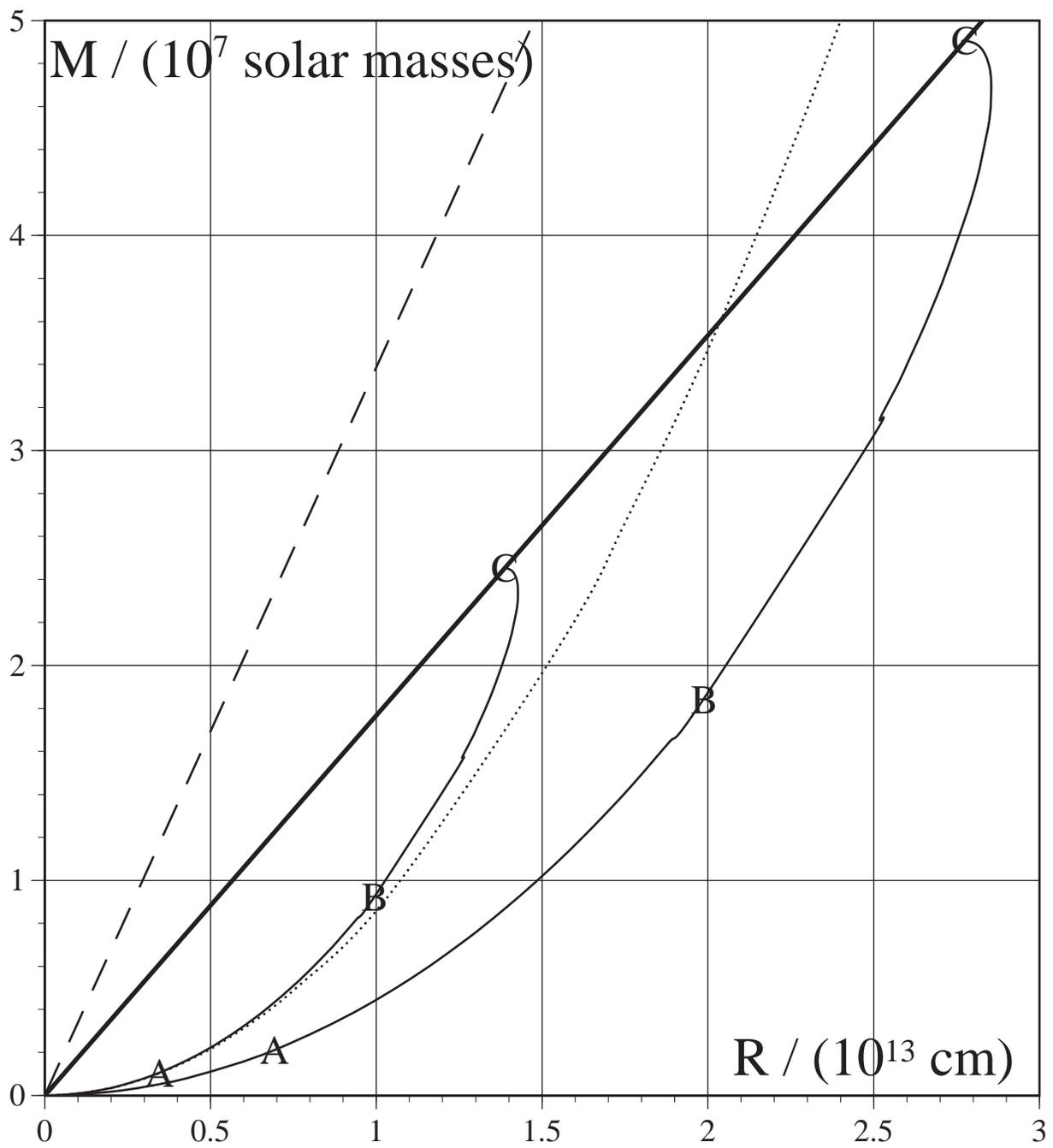
\noindent Figure 1 : The curved solid lines give the balloon mass $M$
versus the balloon radius $R$ for the two values of the wall mass density
$\sigma=1$ TeV${}^3$ and $\sigma=1/2$ TeV${}^3$, while the dotted line
illustrates $M$ versus $R$ in the absence of gravity.  The heavy solid line
gives the maximum mass balloon as $\sigma$ is varied, to be compared
with the dashed line which is $2GM/R=1$.  The energy density profiles
for the points marked $A$, $B$, and $C$ are given in Figure 2.
\vspace{3ex}

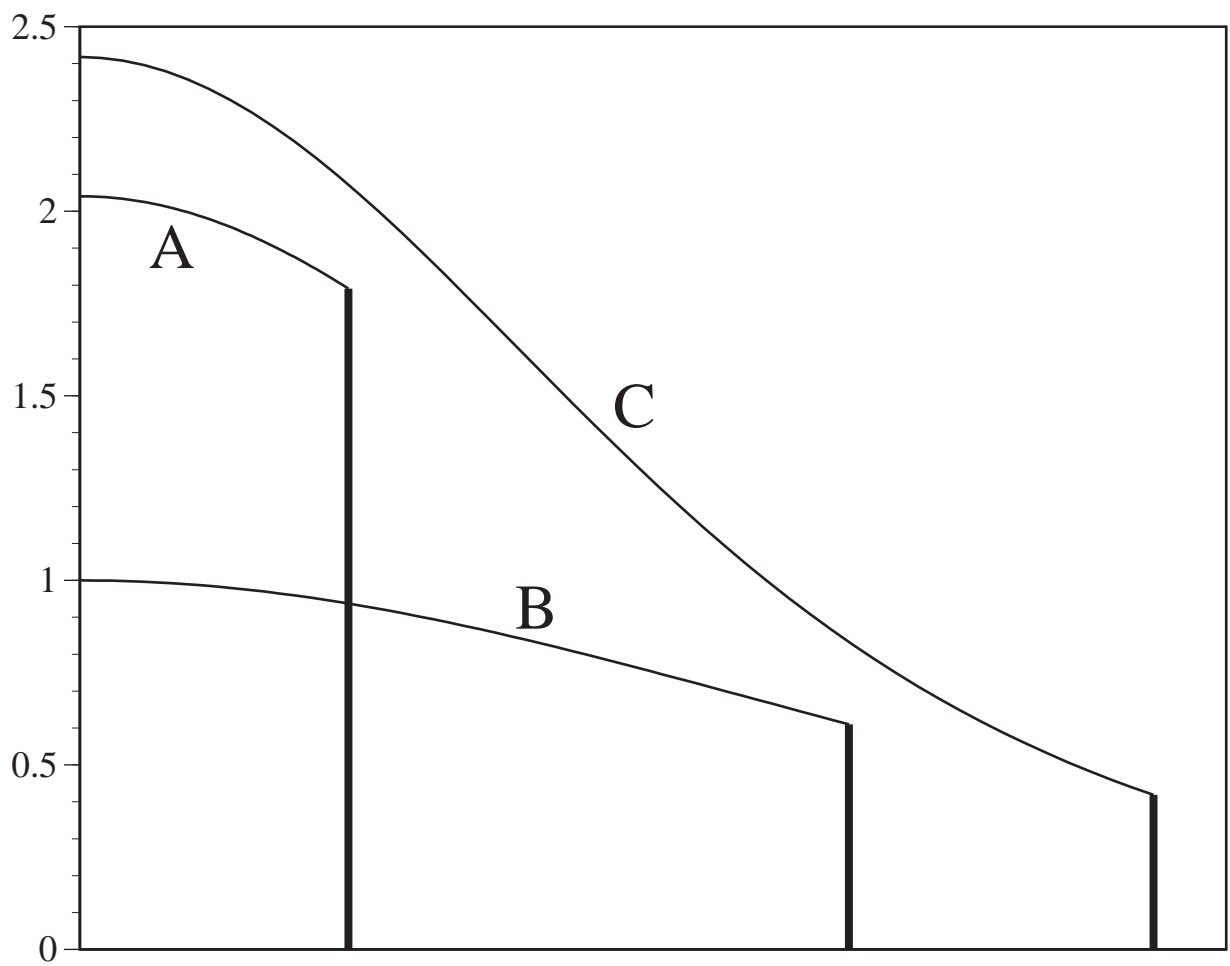
\noindent Figure 2 : The energy density profiles $\rho(r)$ are given for
the points marked by $A$, $B$, and $C$ in Figure 1.  They are given in
units of the smallest possible $\rho(0)$ for a given $\sigma$.


\newpage
\begin{thebibliography}{99}
\bibitem{a} B. Holdom, {\it Phys. Rev. D} {\bf 36} (1987) 1000.
\bibitem{b} A.D.Dolgov and  O. Yu. Markin, {\it Sov. Phys. JETP} {\bf
71} (1990) 207; A.D.Dolgov and  O. Yu. Markin, {\it Prog. Theor. Phys.}
{\bf 85} (1991) 1091. \bibitem{c} B. Holdom and R. A. Malaney, CITA
preprint CITA/93/22 (1993), astro-ph/9306014 (to appear in Ap. J. Lett.).
\bibitem{f} R. Manka, I. Bednarek, D. Karczewska, Silesia University preprint
USL-TH-93-01 (1993), astro-ph/9304007. \bibitem{d} eg. S. Weinberg, {\it
Gravitation and Cosmology}, John Wiley and Sons, 1972. \bibitem{g} W.
Israel, Nuovo Cimento {\bf 44B} (1966) 1. \bibitem{h} J. Ipser and P.
Sikivie, Phys. Rev. {\bf D30} (1984) 712.  \bibitem{e} Waterloo Maple
Software, 450 Philip Street, Waterloo, Ontario, Canada N2L 5J2.
\end{thebibliography}
\end{document}